\documentclass[twocolumn,aps,superscriptaddress, showkeys, nofootinbib,floatfix,linenumbers]{revtex4}
\usepackage{epsfig,bm,feynmf}
\usepackage{graphics}
\usepackage{amsmath}
\usepackage{mathrsfs}
\usepackage{appendix}
\usepackage{bm}
\usepackage[normalem]{ulem}  
\usepackage[dvips]{color} 
\newcommand{\com}[1]{{\sf\color[rgb]{0,0,1}{#1}}}

\renewcommand{\sout}{\bgroup \color{red} \ULdepth=-.5ex \ULset}

\begin{document}
\title{Chiral kinetic approach to the chiral magnetic effect in isobaric collisions}

\author{Yifeng Sun}
\email{sunyfphy@physics.tamu.edu}
\affiliation{Cyclotron Institute and Department of Physics and Astronomy, Texas A$\&$M University, College Station, Texas 77843, USA}%

\author{Che Ming Ko}
\email{ko@comp.tamu.edu}
\affiliation{Cyclotron Institute and Department of Physics and Astronomy, Texas A$\&$M University, College Station, Texas 77843, USA}%

\date{\today}

\begin{abstract}
Based on the chiral kinetic approach using quarks and antiquarks from a multiphase transport model as initial conditions, we study the chiral magnetic effect, i.e., the magnetic field induced separation of charged particles in the transverse plane, in non-central isobaric collisions of Zr$+$Zr and Ru$+$Ru, which have the same atomic number but different proton numbers.  For the observable $\gamma^{OS}-\gamma^{SS}$ related to the difference between the correlations of particles of opposite charges and of same charges, we find a difference between the two collision systems if the magnetic field has a long lifetime of 0.6 fm$/c$ and the observable is evaluated using the initial reaction plane.  This signal of the chiral magnetic effect becomes smaller and comparable to the background contributions from  elliptic flow if the event plane determined from particle emission angles is used.  For the other observable given by the $R(\Delta S)$ correlator related to the distribution of average charge separation in a collision, the signal due to the chiral magnetic effect is found to depend less on whether the reaction or event plane is used in the analysis, and their difference between the two isobaric collision systems is thus a more robust observable. 
\end{abstract}
\keywords{Chiral magnetic effect,  chiral kinetic approach, relativistic heavy ion collisions, AMPT}

\maketitle

\section{introduction}

Experiments at the BNL Relativistic Heavy Ion Collider (RHIC)~\cite{Adams:2005dq,Adcox:2004mh,Arsene:2004fa,Back:2004je} and the CERN Large Hadron Collider (LHC)~\cite{Aamodt:2008zz} have provided  unambiguous evidence for the creation of a quark-gluon plasma (QGP) during the early stage of relativistic heavy ion collisions. Because of topological transitions due to the chiral anomaly in QCD~\cite{PhysRevD.30.2212,PhysRevD.36.581,Moore2011,PhysRevD.93.074036}, non-equal numbers of left- and right-handed quarks can be present and result in a nonzero net axial charge in the QGP.  According to Ref.~\cite{KHARZEEV2008227}, in the presence of the magnetic field produced in non-central heavy ion collisions, the finite net axial charge can lead to a separation of positively and negatively charged particles in the transverse plane of the collisions as a result of the vector charge current generated along the direction of the magnetic field, which is called the chiral magnetic effect (CME)~\cite{KHARZEEV2008227,PhysRevD.78.074033,KHARZEEV2010205}.

Although the phenomenon of charge separation has been observed in experiments at RHIC~\cite{PhysRevLett.103.251601,PhysRevC.81.054908,PhysRevLett.110.012301} and LHC~\cite{PhysRevLett.118.122301},
and also confirmed in studies using the anomalous hydrodynamics~\cite{YIN201642,Hirono:2014oda,Jiang:2016wve}, its explanation is still under debate because of the many background effects from resonance decays~\cite{PhysRevC.70.057901,PhysRevC.95.051901}, the transverse momentum conservation~\cite{PhysRevC.83.014905,PhysRevC.84.024909}, and the local charge conservation~\cite{PhysRevC.83.014913} when the elliptic flow $v_2$ is present.  To separate such $v_2$-driven backgrounds from the CME, collisions involving isobaric systems such as Zr+Zr and Ru+Ru, which have the same atomic but different proton numbers, have been proposed~\cite{PhysRevC.94.041901} and planned at RHIC, because of their similar backgrounds but different CME signals due to the different magnetic fields generated in these collisions as a result of different proton numbers. It has been shown in a schematic study~\cite{PhysRevC.94.041901} and also in more complete studies based on the anomalous hydrodynamics~\cite{Shi:2017cpu} as well as the AMPT model with the an assumed initial charge separation~\cite{Deng:2018dut} that the two collision systems have a relative difference in the charge separation of about ten precent. 

Besides the anomalous hydrodynamics~\cite{PhysRevLett.103.191601}, the chiral kinetic approach has also been developed for studying the chiral magnetic and vortical effects in relativistic heavy ion collisions~\cite{PhysRevLett.109.162001,PhysRevLett.109.181602,PhysRevD.87.085016,PhysRevLett.109.232301,PhysRevLett.110.262301,PhysRevD.90.076007,PhysRevC.94.045204,PhysRevD.96.016023,PhysRevD.95.051901,PhysRevD.95.091901,Huang:2018wdl,PhysRevD.96.016016,PhysRevD.97.016004}.  Using this approach, we have recently studied the effects of chiral magnetic and vortical waves as well as the polarization of $\Lambda$ hyperon in relativistic heavy ion collisions
~\cite{PhysRevC.94.045204,PhysRevC.95.034909,HUANG2018177}.  Since the chiral kinetic approach takes into account the non-equilibrium effect and explicitly treats the $v_2$-driven background without making specific assumptions as in the anomalous hydrodynamics, we use it in the present study to investigate the CME in collisions of the isobaric systems of Ru+Ru and Zr+Zr at RHIC energies.

This paper is organized as follows. In the next section, we describe the chiral kinetic equations of motion for quarks and antiquarks and their modified scattering in the presence of a magnetic field. We then introduce in Sec. III the initial conditions and the magnetic field in relativistic heavy ion collisions, which we take from the AMPT model~\cite{PhysRevC.72.064901}, as well as the initial axial charge density in the partonic matter.  In Sec. IV, we show results on the time evolution and transverse momentum dependence of the charge separation in Ru+Ru collisions via the  $\gamma^{OS}-\gamma^{SS}$ correlator, which is the difference between the correlations of particles of opposite charges and of  same charges, and the $R(\Delta S)$ correlator, which is related to the distribution of average charge separation in a collision, and their differences from Zr+Zr collisions.  Finally, a summary is given in Sec. V.

\section{\com{The} chiral kinetic approach}

In this section, we briefly discuss the chiral kinetic equations of motion for  spin-1/2 fermions and their modified scatterings in a magnetic field.

\subsection{Chiral kinetic equations}

For a massless spin-1/2 quark or antiquark of charge Q and helicity $\lambda$ in a magnetic field ${\bf B}$, the chiral kinetic equations of motion for the rate of changes in its position ${\bf r}$ and momentum ${\bf p}$ are given by~\cite{PhysRevLett.109.162001,PhysRevLett.110.262301,PhysRevC.94.045204}
\begin{eqnarray} 
&&\dot{\mathbf{r}}=\frac{\hat{\mathbf{p}}+Q\lambda(\hat{\mathbf{p}}\cdot\mathbf{b})\mathbf{B}}{1+Q\lambda\mathbf{b}\cdot\mathbf{B}}\label{CKM}\\
&&\dot{\mathbf{p}}=\frac{Q\hat{\mathbf{p}}\times\mathbf{B}}{1+Q\lambda\mathbf{b}\cdot\mathbf{B}}
\label{LF}
\end{eqnarray}
where $\hat{\bf p}$ is a unit vector in the direction of ${\bf p}$ and $\mathbf{b}=\frac{\mathbf{p}}{2p^3}$ is the Berry curvature resulting from the requirement that the spin of the parton follows the direction of its momentum instaneously. To take into account the small $u$ and $d$ quarks masses ($m_u=3$ MeV and $m_d=6$ MeV)~\cite{Olive:2016xmw}, we replace $\hat{\mathbf{p}}$ and $\mathbf{b}$ as $\frac{\mathbf{p}}{E_p}$ and $\frac{\hat{\bf{p}}}{2E_p^2}$ as in Ref.~\cite{PhysRevD.89.094003}. 

\subsection{Parton scattering}

Because of the denominator $\sqrt{G}=1+Q\lambda\mathbf{b}\cdot\mathbf{B}$ in the chiral kinetic equations of motion, the phase-space distribution of partons needs to be multiplied by $\sqrt{G}$~\cite{PhysRevLett.95.137204} to ensure the conservation of vector charge. To obtain the corresponding modified equilibrium distribution of partons from their scatterings, we use the method in our previous study on the $\Lambda$ hyperon polarization in the chiral kinetic approach~\cite{PhysRevC.96.024906}. Specifically, we use the total parton cross section $\sigma_{\rm tot}$ to determine if two partons would collide, as usually used in the parton cascade, but determine their momenta $\mathbf{p}_3$ and $\mathbf{p}_4$ after the scattering with the probability $\sqrt{G(\mathbf{p}_3)}\sqrt{G(\mathbf{p}_4)}$.  As shown in Ref.~\cite{PhysRevC.96.024906}, this automatically leads to the modified equilibrium distribution of partons and the conservation of their vector charge current.

For the parton scattering cross section, we choose it to reproduce the small shear viscosity to entropy density ratio $\eta/s$ in QGP extracted from  experimentally measured anisotropic flows in relativistic heavy ion collisions based on viscous hydrodynamics~\cite{PhysRevLett.99.172301,PhysRevC.78.024902} and transport models~\cite{FERINI2009325,PhysRevC.79.014904}. This empirically determined value is close to the conjectured lower bound for a strongly coupled system in conformal field theory~\cite{PhysRevLett.94.111601} and the values from lattice QCD calculations~\cite{Borsanyi:2013bia}.  For partonic matter dominated by light quarks as considered here, we can relate $\eta/s$ to the total cross section $\sigma_{\rm tot}$ by $\eta/s=\frac{1}{15}\langle p\rangle \tau=\frac{\langle p\rangle}{10n\sigma_{\rm tot}}$~\cite{PhysRevD.90.114009} if the cross section is taken to be isotropic, where $\tau$ is the relaxation time of the partonic matter, $n$ is the parton number density, and $\langle p\rangle$ is the average momentum of partons. Taking $\eta/s=1.5/4\pi$ as determined in Ref.~\cite{PhysRevC.85.024901} from anisotropic flows in relativistic heavy ion collisions using viscous hydrodynamics, we then calculate the parton scattering cross section as a function of parton density and temperature or energy density.

Using the chiral kinetic equations of motion and the above parton scattering cross section, the partonic matter is then evolved until its energy density decreases to $\epsilon_0=0.56$ GeV$/$fm$^3$, similar to the critical energy density from LQCD for the partonic to hadronic transition~\cite{BORSANYI201499} and also that corresponding to the switching temperature $T_{\rm SW}=$ 165 MeV from partonic to the hadronic phases used in viscous hydrodynamics~\cite{PhysRevC.83.054912}.

\section{Initial conditions}

To use the chiral kinetic approach described in the previous section for relativistic  heavy ion collisions, we need information on the initial phase-space distributions of quarks and antiquarks, the initial axial charge density  of produced quark matter, and the time evolution of produced magnetic field. These are discussed in this section. 

\subsection{The AMPT model}

For the initial phase-space distributions of partons, we take them from the string melting version of the AMPT model~\cite{PhysRevC.72.064901} with the values $a=0.5$ and $b=0.9$ GeV$^2$ in the Lund string fragmentation function to give a better description of the charged particle multiplicity density, momentum spectrum, and two- and three-particle correlations~\cite{PhysRevC.84.014903,SUN2017219} in heavy ion collisions at RHIC, as in our previous study on the $\Lambda$ hyperon polarization~\cite{PhysRevC.96.024906}. 

\subsection{Time evolution of  magnetic field}

For the magnetic field produced in non-central heavy ion collisions, it can be calculated from the Lienard-Wiechert potentials. In this potential, the electromagnetic field at a given space-time point ($t,\mathbf{x}$) due to the motion of a charged particle at a constant velocity $\bf v$ is determined by its position ${\bf x}^\prime$ at an earlier time $t^{\prime}$. Using the relation
\begin{eqnarray} 
t-t^{\prime}=|\mathbf{x}-\mathbf{x}^{\prime}|=|\mathbf{x}-(\mathbf{x}_0+\mathbf{v}(t^{\prime}-t_0))|,
\end{eqnarray}
where $\mathbf{x}_0$ is the position of the charged particle at the initial time $t_0$, the magnetic field due to a proton  in the colliding nuclei is then given by
\begin{eqnarray} 
e\mathbf{B}(t,\mathbf{x})=\alpha \frac{(1-v^2)\mathbf{v}\times(\mathbf{x}-\mathbf{x}^{\prime})}{(|\mathbf{x}-\mathbf{x}^{\prime}|-(\mathbf{x}-\mathbf{x}^{\prime})\cdot\mathbf{v})^3},
\end{eqnarray}
where $\alpha$ is the fine-structure constant. 

\begin{figure}[h]
\centering
\includegraphics[width=1\linewidth] {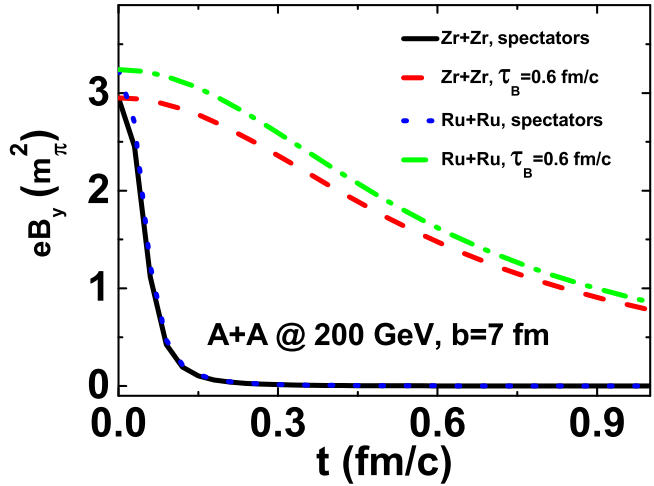}
\caption{(Color online) Time evolution of  magnetic field in the $y$ direction, which is perpendicular to the reaction plane, from  spectator protons as well as from both spectator protons and the QGP with a lifetime $\tau_{B}=0.6$ fm$/c$ for Zr$+$Zr and Ru$+$Ru collisions at $\sqrt{s_{NN}}=200$ GeV and impact parameter $b=7$ fm.}
\label{mag}
\end{figure}

Based on the spatial and momentum information on the protons in the colliding nuclei from AMPT, we can calculate the electromagnetic field produced in heavy ion collisions.  As the colliding nuclei move towards each other with the same velocity in opposite directions along the beam direction, taken as the $z$ direction, and with their centers located at $x=\pm b/2$ in the $x$ direction of the reaction plane, where $b$ is the impact parameter of the collision, the magnetic field in the overlap region of the two nuclei is then in the $y$ direction.  Figure~\ref{mag} shows the time evolution of the magnetic field at $\mathbf{x}=0$ in the $y$ direction for the isobaric collisions of Zr$+$Zr and Ru$+$Ru at $\sqrt{s_{NN}}=200$ GeV and impact parameter b=7 fm.  The black solid and blue dotted lines are the time evolution of the magnetic field in Zr$+$Zr and Ru$+$Ru collisions due to spectator protons. The magnetic field is seen to have a very large value of about 2.95 $m_{\pi}^2$ and 3.24 $m_{\pi}^2$ for the two collision systems, respectively, and decreases rapidly in time with a lifetime of about $\tau_{B}=0.066$ fm$/c$ in both cases. We note that there are also studies on the fluctuation of magnetic field  due to the proton position fluctuations in nuclei~\cite{BZDAK2012171,PhysRevC.85.044907}. Since protons in these studies are assumed to be point-like with their positions  randomly determined from a Woods-Saxon (WS) distribution, the obtained magnetic field fluctuation may have been overestimated. 

Because the hot medium created in heavy ion collisions is a conducting plasma, it can in principle increase the lifetime of the fast decaying magnetic field from spectator protons. Studies on this effect have, however, led to very different conclusions~\cite{McLerran2014184,PhysRevC.89.054905,PhysRevC.93.014905,PhysRevC.94.044903}. In the present study, we adopt the magnetic field used in Refs.~\cite{PhysRevC.89.044909,Jiang:2016wve,Shi:2017cpu} with $eB=\frac{eB_0}{1+(\frac{t}{\tau_B})^2}$ and $\tau_B=$ 0.6 fm$/c$ as an upper limit for the lifetime, as shown by the red dashed and green dash-dotted curves in Fig.~\ref{mag} for Zr+Zr and Ru+Ru collisions, respectively, to illustrate the CME in relativistic heavy ion collisions. 

\subsection{Axial charge density}

The other essential input in the study of CME is the axial charge density $n_5$.  Although $n_5$ is usually zero in hot hadronic matter, the QGP produced in relativistic heavy ion collisions can have event-by-event fluctuating net axial charges as a result of its topological transitions due to  sphalerons~\cite{PhysRevD.30.2212,PhysRevD.36.581,Moore2011,PhysRevD.93.074036}. As shown in Refs.~\cite{Hirono:2014oda,KHARZEEV2002298,PhysRevC.82.057902}, the topological transition rate is related to fluctuations of the color electromagnetic field configurations, resulting in  the following axial charge density fluctuation in relativistic heavy ion collisions~\cite{Hirono:2014oda,Jiang:2016wve}: \begin{eqnarray} 
\sqrt{n_5^2}&=&\tau_0\frac{g^2\mathbf{E}^c\cdot\mathbf{B}^c}{16\pi^2}\times\sqrt{N_{\rm tube}}\times \frac{\pi \rho_{\rm tube}^2}{A_{\rm overlap}}\nonumber
\\&\approx&\tau_0\frac{Q_{\rm s}^4}{16\pi^2}\times\sqrt{N_{\rm coll}}\times \frac{\pi \rho_{\rm tube}^2}{A_{\rm overlap}},
\end{eqnarray}
where $\tau_0=0.6$ fm$/c$ is the thermalization time of QGP, $Q_{\rm s}^2\approx$ 1.25 GeV$^2$ is the saturation scale for gluons in the colliding nuclei at RHIC energy of 200 GeV~\cite{PhysRevD.87.034002,PhysRevLett.100.022303}, $N_{\rm tube}$ and $\rho_{\rm tube}\approx 1$ fm are the number and radius of the glasma flux tube, and $A_{\rm overlap}$ is the transverse area of the overlap region of colliding nuclei. The second line in the above equation follows from the approximation $N_{\rm tube}\approx N_{\rm coll}$, where $N_{\rm coll}$ is the number of binary collisions between nucleons in the colliding nuclei and can be determined by using the Glauber Model. With the value $N_{\rm coll}= 82.73$ for Zr$+$Zr and Ru$+$Ru collisions at $b=$ 7 fm, we find that the fluctuation in the topological charge associated with the color electromagnetic field configuration during the initial state of the collision is $\sqrt{N_5^2}=2\tau_0 A_{\rm overlap}\sqrt{n_5^2}=\frac{2\tau_0^2\pi\rho_{\rm tube}^2Q_s^4\sqrt{N_{\rm coll}}}{16\pi^2}=135.16$. This fluctuation can be modeled by letting the helicity of a parton from the AMPT model to have the probability of $p=\frac{1+\sqrt{N_5^2}/N}{2}$, where $N$ is total number of partons before $\tau_0$ in an event, to be positive in half of the events and to be negative in the other half of the events.  

\section{Results}

With above initial parton distributions and the net axial charge density for collisions at $\sqrt{s_{NN}}=200$ GeV and impact parameter b=7 fm, we first consider Ru+Ru collisions by letting all quarks follow the chiral kinetic equations of motion and collide with each other with modified scatterings in the presence of an external magnetic field in $y$ direction until the critical energy density of 0.56 GeV$/$fm$^3$ used in the hydrodynamic approach~\cite{PhysRevC.83.054912}. We then study the charge separation of light quarks at mid-pseduorapidity ($|\eta|\le$1). Similar calculations are  carried out for Zr+Zr collisions to see how the results differ from those for the Ru+Ru collisions.

\subsection{Charge separation in Ru$+$Ru collisions}

\begin{figure}[h]
\centering
\includegraphics[width=1\linewidth] {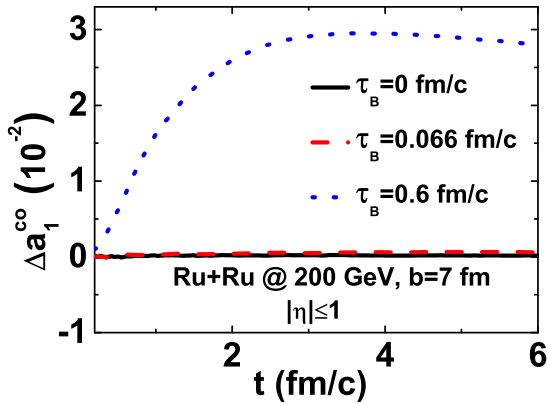}
\caption{(Color online) Time evolution of the vector charge dipole moment of mid-pseudorapidity ($|\eta|\le1$) light quarks in coordinate space in  Ru$+$Ru collisions at $\sqrt{s_{NN}}=200$ GeV and impact parameter b=7 fm for different magnetic field lifetimes. }
\label{a1c}
\end{figure}

We first show in Fig.~\ref{a1c} the time evolution of the vector charge dipole moment $\Delta a_1^{\rm co}=\langle \sin\phi_+
\rangle-\langle \sin\phi_-\rangle$ of  mid-pseudorapidity ($|\eta|\le$1) light quarks in coordinate space in Ru$+$Ru collisions for different  magnetic field lifetimes in  events with more right-handed than left-handed quarks.  In the above, $\phi_+$ and $\phi_-$ are  azimuthal angles of the position vectors of positively and negatively charged quarks in the transverse plane of a collision, and the average is over all light quarks from these events.  The black solid and red dashed lines are, respectively, the results obtained without the magnetic field and with a short-lived magnetic field from  spectator protons ($\tau_B=$ 0.066 fm$/c$). In both cases, the resulting vector charge dipole moment is essentially zero at all times.  With a lifetime of 0.6 fm$/c$ for the magnetic field, the vector charge dipole moment  increases appreciably with time as a result of the CME and becomes almost constant after $t=3$ fm$/c$ due to the decay of the magnetic field.  We note that for the other half events with more left-handed than right-handed quarks, the vector charge dipole moment has same magnitude but opposite sign compared to the above case. 

\begin{figure}[h]
\centering
\includegraphics[width=1\linewidth] {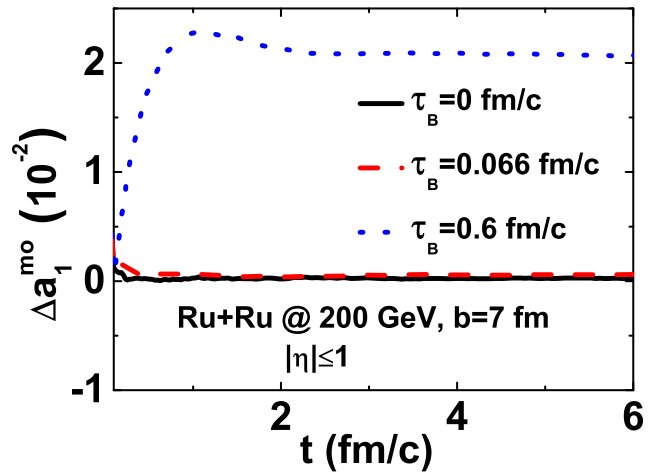}
\caption{(Color online) same as Fig.~\ref{a1c} for the time evolution of the vector charge dipole moment in momentum space.}
\label{a1p}
\end{figure}

In Fig.~\ref{a1p}, we plot the vector charge dipole moment in momentum space $\Delta a_1^{\rm mo}$, which is similarly defined as $\Delta a_1^{\rm co}$ by using the azimuthal angle of  parton momentum in the transverse plane.  As for $\Delta a_1^{\rm co}$, $\Delta a_1^{\rm mo}$ is essentially zero when the magnetic field is absent or has a very short lifetime as shown by the black solid and red dashed lines, respectively, and is large in the presence of a long-lived magnetic field as shown by the blue dotted line.  Although the $\Delta a_1^{\rm mo}$ in the latter case initially increases, it slightly decreases afterwards until it reaches a large constant value. The reason for this behavior of $\Delta a_1^{\rm mo}$ is as follows.  Because of the coupling between the magnetic field and the spin of charged quarks,  positively charged quarks are quickly polarized along the positive $y$ direction while negatively charged quarks are polarized along the negative $y$ direction by the modified collisions. Since the number of quarks and anti-quarks of positive helicity is larger than that of negative helicity in these events, positively charged quarks acquire a net momentum in the positive $y$ direction while the net momentum of negatively charged quarks is in the negative $y$ direction, resulting in a fast increase of $\Delta a_1^{\rm mo}$ during  early times.  As the magnetic field decays with time, $\Delta a_1^{\rm mo}$ then decreases with the decreasing spin polarization, which is, however, compensated by the large increase of  $\Delta a_1^{\rm co}$, leading thus to a large and constant $\Delta a_1^{\rm mo}$. 

\subsection{Transverse momentum dependence of charge separation}

\begin{figure}[h]
\centering
\includegraphics[width=0.9\linewidth] {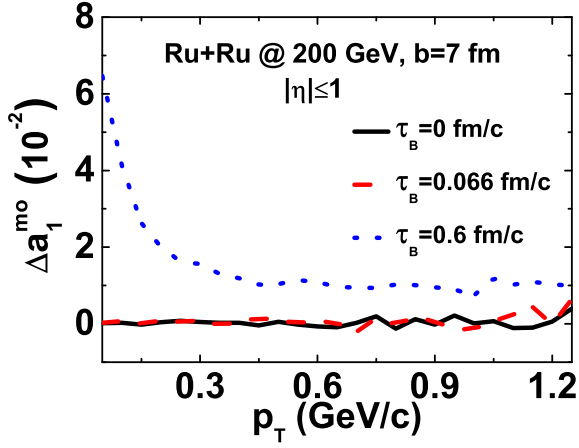}
\caption{(Color online) Same as Fig.~\ref{a1c} for the transverse momentum dependence of $\Delta a_1^{\rm mo}$ of light quarks for different magnetic field lifetimes.}
\label{pt}
\end{figure}

Results for the transverse momentum dependence of vector charge dipole moment in momentum space or separation $\Delta a_1^{\rm mo}$ of light quarks are shown in Fig.~\ref{pt}, again from events with more right- than left-handed quarks. In both cases of without magnetic field and a strong magnetic field with a short lifetime, the charge separation is negligible for quarks of any momentum as shown by black solid and red dashed lines in Fig.~\ref{pt}.  In the presence of a strong and long-lived magnetic field,  a large charge separation appears as shown by the blue dotted line in Fig.~\ref{pt}.  Although the value of the charge separation is comparable to that from the anomalous-viscous fluid dynamics (AVFD)~\cite{Shi:2017cpu}, its transverse momentum dependence is different due to the stronger CME in the chiral kinetic approach for quarks of low momentum as a result of the Berry curvature $\frac{\mathbf{p}}{2p^3}$ in the equations of motion and the modified phase-space distribution.

\subsection{The $\gamma^{\rm OS}-\gamma^{\rm SS}$ correlator}

The CME-driven charge separation can lead to different azimuthal distributions for positively and negatively charged particles in an event, given by 
\begin{eqnarray} 
\frac{dN^{\pm}}{d\phi}\propto 1+2v_2\cos(2\phi-2\Psi_{\rm RP})\pm2a_1\sin(\phi-\Psi_{\rm RP}),
\label{a1}
\end{eqnarray}
where $\Psi_{\rm RP}$ is the initial reaction plane of a collision. Since the topological charge of the partonic matter has the same probability for being positive and negative, $\langle a_1\rangle=0$, to measure the charge separation signal in experiments is usually through the $\gamma^{\rm OS}-\gamma^{\rm SS}$ correlator, which is the difference between the correlations of particles of opposite signs and of same signs in charges, defined, respectively, by
\begin{eqnarray} 
&&\gamma^{\rm OS}=\left \langle \cos(\phi_{\alpha}^{+(-)}+\phi_{\beta}^{-(+)}-2\Psi_{\rm RP}) \right \rangle,\nonumber
\\&&\gamma^{\rm SS}=\left \langle \cos(\phi_{\alpha}^{+(-)}+\phi_{\beta}^{+(-)}-2\Psi_{\rm RP}) \right \rangle,
\end{eqnarray}
where $\phi_{\alpha}$ and $\phi_{\beta}$ are the azimuthal angles for same-sign ($++$ or $--$) and opposite-sign ($+-$) charged particle pairs, respectively. We note that if the charges of all particles are randomly chosen, then one has from Eq.~(\ref{a1}) $\gamma^{\rm OS}-\gamma^{\rm SS}=2a_1^2=(\Delta a_1^{\rm mo})^2/2$.

\begin{figure}[h]
\centering
\includegraphics[width=0.9\linewidth] {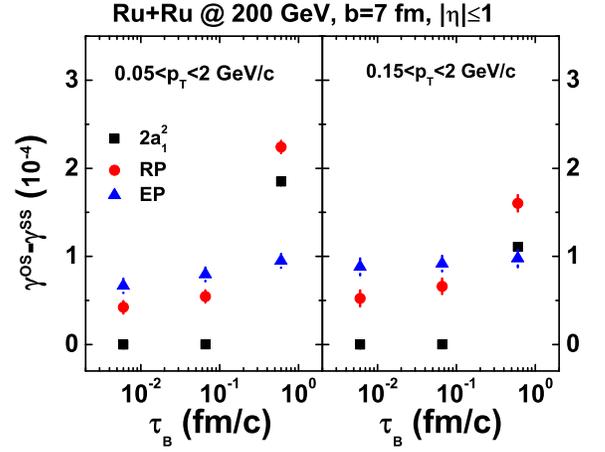}
\caption{(Color online) Magnetic field lifetime dependence of the $\gamma^{OS}-\gamma^{SS}$ correlator of of mid-pseudorapidity ($|\eta|\le1$) light quarks in Ru$+$Ru collisions at $\sqrt{s_{NN}}=200$ GeV and impact parameter b=7 fm for different transverse momentum ranges and using different calculation methods. The error bars denote the statistical errors due to the finite numberer of events used in the study.}
\label{Ru}
\end{figure}

In the left panel of Fig.~\ref{Ru}, we show the $\gamma^{OS}-\gamma^{SS}$ correlator of light quarks in the transverse momentum range $0.05<p_T<2.0 $ GeV$/c$ for Ru+Ru collisions at $\sqrt{s_{NN}}=200$ GeV and  b=7 fm from all events with either more right-handed quarks or left-handed quarks. The black squares denote its value calculated directly from $2a_1^2$, which is almost zero when the lifetime $\tau_B$ of the magnetic field is zero or 0.066 fm$/c$, and is 1.849$\times 10^{-4}$ when the lifetime is 0.6 fm$/c$. Taking into consideration of event by event fluctuations by using the theoretical reaction plane $\Psi_{\rm RP}=0$ from the AMPT model, all results are increased by a factor of about 0.5 $\times 10^{-4}$,  shown by the red circles with error bars due to the finite number of events, indicating that there is a $v_2$-driven background in the AMPT model during the partonic phase.

Using the event plane reconstructed from emitted particles as introduced in Ref.~\cite{PhysRevC.58.1671}, that is
\begin{eqnarray} 
&&\Psi_{\rm EP}=\frac{1}{2}\tan^{-1} \frac{\sum_i \omega_i \sin(2\phi_i)}{\sum_i \omega_i \cos(2\phi_i)},\end{eqnarray}
where the summation is over all particles in the phase-space cut in an event and taking $\omega_i=p_{iT}$, the resulting $\gamma^{OS}-\gamma^{SS}$ correlator is shown by the blue triangles with error bars in Fig.~\ref{Ru}. Its value shows a weaker dependence on the lifetime of the magnetic field and has a value of about 0.8$\times 10^{-4}$. This is due to the smaller correlation between the event plane and the initial reaction plane in these small systems, where $\left \langle \cos(2\Psi_{\rm EP}-2\Psi_{\rm RP})  \right \rangle=0.27$.

For the transverse momentum range $0.15<p_T<2.0 $ GeV$/c$, the results are similar, although the value of the correlator decreases by 0.7$\times 10^{-4}$ either using the initial reaction plane  or  the value of $2a_1^2$ in the calculation if there is a long-lived magnetic field, as a result of the smaller CME effect on quarks of higher momentum in the chiral kinetic approach. Using the event plane from emitted particles still shows a weak dependence on the lifetime of the magnetic field, even though the value slightly increases compared to the case of including quarks of lower momentum. 

\subsection{The $R(\Delta S)$ correlator}

Recently, a new correlator $R(\Delta S)$ has been proposed to measure the strength of CME~\cite{Magdy:2017yje}, which is defined as 
\begin{eqnarray} 
R(\Delta S)=\frac{C(\Delta S)}{C^{\perp }(\Delta S)},
\end{eqnarray}
where 
\begin{eqnarray} 
&&C(\Delta S)=\frac{N_{\rm real}(\Delta S)}{N_{\rm shuffled}(\Delta S)},\\
&&C^\perp(\Delta S)=\frac{N_{\rm real}^\perp(\Delta S)}{N_{\rm shuffled}^\perp(\Delta S)}.
\end{eqnarray}
In the above, $N_{\rm real}(\Delta S)$ is the distribution of average charge separation in each event, which is defined as~\cite{PhysRevC.83.011901}
\begin{eqnarray} 
&&\Delta S=s^{+}-s^{-}\nonumber
\\&&s^{+}=\frac{\sum_i \sin(\phi_i^{+}-\Psi_{\rm RP})}{N_{\rm p}}\nonumber
\\&&s^{-}=\frac{\sum_i \sin(\phi_i^{-}-\Psi_{\rm RP})}{N_{\rm n}},
\end{eqnarray}
where the summation is over the numbers $N_p$ and $N_n$  of positively and negatively charged particles in an event. The $N_{\rm shuffled}(\Delta S)$ is the distribution obtained by randomly choosing $N_p$ particles from $N_p+N_n$ charged particles in this event and setting them as positively charged particles and the rest as negatively charged particles. $N_{\rm real} ^{\perp }(\Delta S)$ and $N_{\rm shuffled} ^{\perp }(\Delta S)$ are similarly calculated by using the same procedure after replacing $\Psi_{\rm RP}$ by $\Psi_{\rm RP}+\pi/2$.

\begin{figure}[h]
\centering
\includegraphics[width=0.9\linewidth] {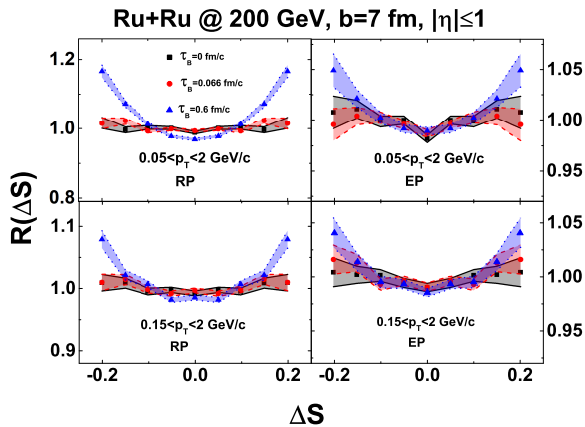}
\caption{(Color online) $\Delta S$ dependence of the $R(\Delta S)$ correlator of mid-pseudorapidity ($|\eta|\le1$) light quarks in Ru$+$Ru collisions at $\sqrt{s_{NN}}=200$ GeV and impact parameter b=7 fm for different lifetimes of the magnetic field, transverse momentum ranges, and reaction planes. Widths of colored bands correspond to the statistical errors due to the finite number of events used in the study.}
\label{rur}
\end{figure}

If the charges of all particles are randomly selected according to Eq.~(\ref{a1}), then all four $\Delta S$'s have zero average values and the variances 
\begin{eqnarray}
\left \langle (\Delta S_{\rm real})^2 \right \rangle&=&\frac{1-v_2/2}{2N_{\rm p}}+\frac{1-v_2/2}{2N_{\rm n}}\nonumber\\
&+&\left(4-\frac{1}{N_p}-\frac{1}{N_n}\right)a_1^2,\\
\left \langle (\Delta S_{\rm shuffled})^2 \right \rangle&=&\frac{1-v_2/2}{2N_{\rm p}}+\frac{1-v_2/2}{2N_{\rm n}}\nonumber\\
&+&\left(\frac{4}{N_p+N_n-1}-\frac{1}{N_p}-\frac{1}{N_n}\right)a_1^2,\nonumber\\
\\
\left \langle (\Delta S_{\rm real}^{\perp })^2\right \rangle&=&\frac{1+v_2/2}{2N_{\rm p}}+\frac{1+v_2/2}{2N_{\rm n}},\\
\left \langle (\Delta S_{\rm shuffled}^{\perp })^2 \right\rangle&=&\frac{1+v_2/2}{2N_{\rm p}}+\frac{1+v_2/2}{2N_{\rm n}}.
\end{eqnarray}
In this case, the $R(\Delta S)$ correlator depends on both the selected charged particle multiplicities $N_{\rm p}$ and $N_{\rm n}$ as well as the charge separation signal $a_1^2$. 

Results on the $\Delta S$ dependence of the $R(\Delta S)$ correlator calculated from light quarks at the critical energy density and based on both events with more and less right-handed than left-handed quarks are shown in Fig.~\ref{rur} by colored bands to include the statistical errors due to the finite number of events used in the study.  The left upper panel of Fig.~\ref{rur} shows the $R(\Delta S)$ correlator calculated by using the initial reaction plane of the collision for light quarks in the transverse momentum range of $0.05<p_T<2$ GeV$/c$. Similar $R(\Delta S)$ correlators are obtained for the cases without magnetic field and in the presence of a short-lived magnetic field, and both slightly increase at large $|\Delta S|$. The behavior of $R(\Delta S)$ changes significantly in the presence of a long-lived magnetic field, particularly the large increase in its value at large $|\Delta S|$, which is similar to the results from Ref.~\cite{Magdy:2018lwk} based on the AVFD approach. 

Under the same conditions as above but using the event plane of light quarks, we see from the right upper panel of Fig.~\ref{rur} that the differences in the behaviors of $R(\Delta S)$ between the different scenarios for the lifetime of the magnetic field is still appreciable, although the value of $R(\Delta S)$ correlator at large  $|\Delta S|$ in the presence of a long-lived magnetic field is less pronounced.

Because the decrease of $a_1^2$ and the increase of $1/N_{\rm p}$ and $1/N_{\rm n}$ for light quarks in the momentum range $0.15<p_T<2$ GeV$/c$, we can conclude that the $R(\Delta S)$ correlator should show a less concave shape than for light quarks in the momentum range $0.05<p_T<2$ GeV$/c$, as shown in the lower panels of Fig.~\ref{rur}. However, the $R(\Delta S)$ correlator for the case of a long-lived magnetic field is still different from that of a short-lived magnetic field whether one uses the initial reaction plane or the event plane from emitted particles, implying that the $R$ correlator is a more robust signal than the $\gamma^{\rm OS}-\gamma^{\rm SS}$ correlator for the CME.

\subsection{Comparison between isobaric Zr$+$Zr and Ru$+$Ru collisions}

As the signal of CME can be contaminated by the $v_2$-driven background, there are suggestions that isobaric collisions of Zr$+$Zr and Ru$+$Ru can be used to separate the background from the CME~\cite{PhysRevC.94.041901}. In this section, we compare results from the chiral kinetic approach for these two collision systems to study  the CME signal.

\begin{figure}[h]
\centering
\includegraphics[width=1\linewidth] {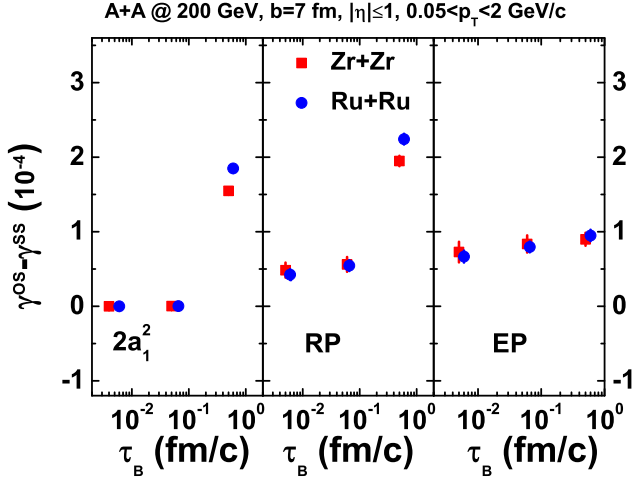}
\caption{(Color online) Same as Fig.~\ref{Ru} for mid-pseudorapidity light quarks in the transverse momentum range $0.05<p_T<2$ GeV$/c$  for Zr+Zr and Ru+Ru collisions.}
\label{gamma}
\end{figure}

Shown in Fig.~\ref{gamma} are the results for the $\gamma^{OS}-\gamma^{SS}$ correlator of mid-pseudorapidity light quarks in the transverse momentum range $0.05<p_T<2$ GeV$/c$ from Zr+Zr and Ru+Ru collisions. It is seen that there are almost no difference between the results from these two collision systems if the lifetime of the magnetic field is short. In  the presence of a long-lived magnetic field, the charge separation $2a_1^2$ due to the CME changes from 1.549$\times 10^{-4}$ for Zr+Zr collisions to 1.849$\times 10^{-4}$ or Ru+Ru collisions, which shows a 19.4\% increase in these two collision systems and is consistent with the results from from Ref.~\cite{Shi:2017cpu} based on the AVFD approach. Using the initial reaction plane by calculating the $\gamma^{OS}-\gamma^{SS}$ correlator event-by-event, the charge separation changes from 1.949$\times 10^{-4}$ to 2.242$\times 10^{-4}$, which increases by about 15.0$\pm$5.4\% and is still appreciable.  Using the event plane reconstructed from azimuthal angles of emitted particles, the two collision systems show, however, a negligible difference.  Because of the difficulty in determining the initial reaction plane in experiments, to extract the CME using the $\gamma^{OS}-\gamma^{SS}$ correlator is thus not an easy task. 
 
\begin{figure}[h]
\centering
\includegraphics[width=1\linewidth] {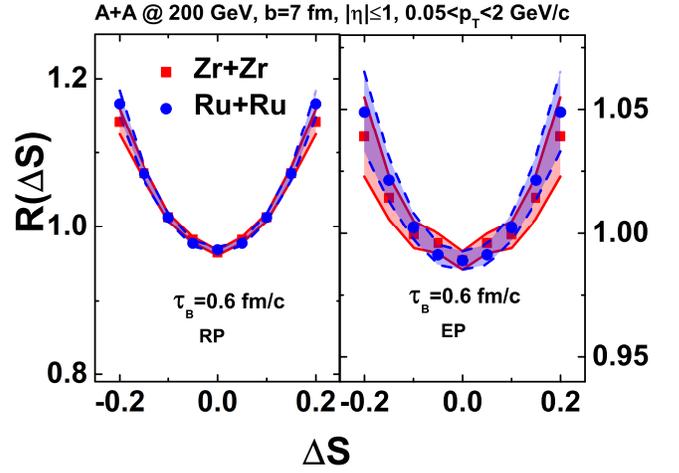}
\caption{(Color online) Same as Fig.~\ref{rur} for mid-pseudorapidity light quarks of transverse momenta $0.05<p_T<2$ GeV$/c$ in Zr+Zr and 
Ru+Ru collisions in the presence of a magnetic field of long lifetime ($\tau_B=$ 0.6 fm$/c$).}
\label{rr}
\end{figure}

Since the $R(\Delta S)$ correlators for the two collision systems do not show clear concave shapes if the lifetime of magnetic field is short, we compare the two collision systems in the presence of a long-lived magnetic field. As shown in Fig.~\ref{rr}, the $R(\Delta S)$ correlator indeed have concave shapes in both Zr+Zr and Ru+Ru collisions.  Because of the small difference in the produced magnetic field in these two collision systems, the difference in their $R(\Delta S)$ correlators is smaller than the statistical errors in the present study as a result of the finite number of events used in our calculations, and this is the case whether the initial reaction plane or the event plane from emitted particles is used.  On the other hand, it has been shown in Ref.~\cite{Magdy:2018lwk} that with more events included in the study, it is possible to extract the signal of CME from the $R(\Delta S)$ correlator, particularly at large $\Delta S$.  Also,  the difference in the $R(\Delta S)$ correlator with and without a long-lived magnetic field is large, which makes the $R(\Delta S)$ correlator a plausible observable for identifying the signal of CME. 

\section{Summary}

Using the chiral kinetic approach based on initial conditions, including the net axial charge density, taken from the AMPT model, we have studied the charge separation signal of light quarks in Zr+Zr and Ru+Ru collisions in the presence of a magnetic field. We have found that these two isobaric collision systems have large charge separation signals ($\Delta a_1^{\rm mo.}$) due to the CME in the presence of a long-lived magnetic field, which becomes, however, negligible if the lifetime of magnetic field is short.  By studying both the $\gamma^{OS}-\gamma^{SS}$ and $R(\Delta S)$ correlators of light quarks in mid-pseudorapidity, we have further found that  without the magnetic field or in the presence of a short-lived magnetic field from spectator protons ($\tau_B$=0.066 fm$/c$), there are almost no differences in Zr+Zr and Ru+Ru collisions for both two correlators, indicating that the two collision systems are non-distinguishable if the lifetime of the magnetic field is short. In the presence of a long-lived magnetic field ($\tau_B=$ 0.6 fm$/c$), the two collision systems are found to show similar behaviors in the $R(\Delta S)$ correlator because of the very similar magnetic fields produced in their collisions, but both are different from that of without the magnetic field or a short-lived magnetic field. These results are independent of whether the initial reaction plane or event plane from emitted particles is used in the analysis.  For the $\gamma^{OS}-\gamma^{SS}$ correlator, these two collision systems show a difference of about 15\% if the theoretical reaction plane is used.  Because of the difficulty in determining the initial reaction plane in experiments, using the event plane from emitted particles will lead to a negligible difference in the CME signals between these two collision systems as a result of  the small correlation between the two reaction planes in small systems, thus making it hard to observe the CME from the $\gamma^{OS}-\gamma^{SS}$ correlator. 

Our results are obtained without including the effect from the hadronic stage of relativistic heavy ion collisions, which could lead to background effects due to resonance decays and local charge conservation on charge separation~\cite{Bozek:2017plp,Feng:2018chm}.  Since these effects are similar in collisions of isobaric systems, they are not expected to affect the conclusion from the present study. 

\section*{ACKNOWLEDGEMENTS}

This work was supported in part by the US Department of Energy under Contract No. DE-SC0015266 and the Welch Foundation under Grant No. A-1358.

\bibliography{ref}

\end{document}